\documentstyle[12pt]{article}
\newcommand{\beq}{\begin{equation}}
\newcommand{\eeq}{\end{equation}}

\begin{document}
\def\lag{\langle}
\def\rag{\rangle}
\begin{titlepage}
\begin{flushright}
Bielefeld, BI-TP 91/21
\end{flushright}
\vspace{0.5cm}
\begin{center}
\large\bf{ Multicanonical Potts Model Simulation$^{1}$}\\
\vspace{1.5cm}
\large{
Bernd A. Berg$^{2,3,4}$
and Thomas Neuhaus$^2$
}
\end{center}
\vspace{3cm}
\begin{center}
{\bf Abstract:}
\end{center}

Relying on the recently proposed multicanonical algorithm, we present
a numerical simulation of the first order phase transition  in
the $2d$ 10-state Potts model on lattices up to sizes $100 \times 100$.
It is demonstrated that the new algorithm {\it lacks} an exponentially
fast increase of the tunneling time between metastable states
as a function of the linear size $L$ of the system. Instead, the
tunneling time diverges approximately proportional to $L^{2.65}$.
Thus the computational
effort as counted per degree of freedom
for generating an independent configuration in
the unstable region of the model
rises proportional to $V^{2.3}$, where
$V$ is the volume
of the system.
On our largest lattice we gain more than two orders of magnitude as
compared to a standard heat bath algorithm. As a first physical
application we report a high precision computation of the interfacial
tension.

\vfill
\footnotetext[1]{{This
research project was partially funded by the National
Science Foundation under grant INT-8922411 and by the the Department
of Energy under contract DE-FG05-87ER40319}}
\footnotetext[2]{{\em
 Fakult\"at f\"ur Physik, Universit\"at Bielefeld,
 D-4800 Bielefeld, FRG}}
\footnotetext[3]{{\em
 Supercomputer Computations Research Institute Tallahassee, FL
 32306, USA}}
\footnotetext[4]{{
On leave of absence from Department of Physics, The Florida State University,
Tallahassee, USA. }}
\end{titlepage}

Critical slowing down is of crucial importance to computer simulations 
of phase transitions. For second order phase transitions long
autocorrelation times at criticality cause severe restrictions on the
maximum lattice size for which one can obtain good statistics
of thermodynamic quantities.
For a number of spin systems this critical slowing down was overcome by
the nonlocal cluster algorithm of Swendsen-Wang \cite{SweWa}, for
a recent review see \cite{SweWa2}.
However, for first order transition one encounters an even worse
and different problem of critical slowing down.
The interfacial free energy between disordered and ordered states
has a finite value on the
critical point for the infinite volume system.
Configurations dominated by the presence of the interface
will be exponentially suppressed by the Boltzmann
factor in the canonical ensemble. On  finite lattices this
leads then to an exponentially fast suppression
of the tunneling between metastable states of the system
with increasing lattice size. To overcome
this critical slowing down effect for first order transitions, we recently
proposed a {\it multicanonical} Monte Carlo algorithm \cite{our}.
The multicanonical MC algorithm is designed to enhance
configurations, which are
dominated by the presence
of the interface and therefore exponentially suppressed.
In this way it is possible to avoid the exponentially
fast growing slowing down at the first order phase transition.
In this paper we demonstrate this in the case of our example:
the $2d$ $10$-state Potts model.

The $2d$ $10$-state Potts model \cite{Potts}
is defined by the partition function
\begin{eqnarray}
Z(\beta ) & = & \sum_{\rm configurations} \exp (\beta S)   , \\
S & = &  \sum_{<i,j>} \delta_{q_i,q_j}    , \\
q_i,q_j & = & 0,...,9.
\label{eq:model}
\end{eqnarray}
Recently there has been renewed interest in this model \cite{Bill,Lee,Borgs}.
It serves as an excellent laboratory for finite size scaling (FSS) studies 
of temperature driven strong first order phase transitions.  We have chosen
this model as the first testing ground for our new method by, essentially,
two reasons: a) it has a strong first order transition and this is the
situation for which our method promises the most dramatic improvements,
and b) accurate data, due to recent large scale simulations \cite{Bill,Borgs},
exist in the literature. Beyond these reasons, an accurate determination of
the interfacial free energy is certainly also of physical interest.

To calculate the interfacial free energy $F^s$ between the disordered and 
the (ten) ordered states has remained the hardest problem.
The reason is the pronounced
double peak structure of the sampled action density $P_L(S)$
in the canonical ensemble near the critical
point, as illustrated in Fig. 1 for lattices of size $V=L^2$ with
$L=24$ and $L=100$. The figure relies on data obtained with
the multicanonical MC algorithm
to be discussed in this paper, and is arranged to
correspond to the action density of the canonical ensemble at
$\beta = \beta^c_L$. The pseudocritical point
$\beta^c_L$ is defined such that both maxima
are of equal height:
\begin{equation}
P_L^{1,\max} = P_L(S^{1,max}_L) = P_L(S^{2,max}_L) = P_L^{2,\max}.
\label{eq:pmax}
\end{equation}
In addition we have imposed the normalization condition
$1=P_L^{1,\max}=P_L^{2,\max}$.
Fig. 2  depicts the action
densities for lattices with $L=16,24,34,50,70$ and $L=100$
on a logarithmic scale and we see that $4$ orders of magnitude
are involved: $P_L^{\min}/P_L^{\max} \simeq 5.1 \times 10^{-5}$ for $L=100$.
With our conventions for $P_L(S)$ the interfacial free energy
$F^s=F^s_{\infty}$ can now be defined \cite{face}
as the $L\to\infty$ limit of the quantity
\begin{equation}
F^s_L\ =\ - {1\over L} \ln  P_L^{\min}  .
\label{eq:FS}
\end{equation}
For a numerical calculation of $F^s_L$ it is now clear that any
algorithm which samples configurations with a probability $\sim P_L(S)$
would slow down proportional to $1/P_L^{\min}$. As for large lattices
$P_L^{\min} \sim \exp (-F^s L^{d-1})$, it is expected that an
appropriately defined tunneling time $\tau_L^t$ will behave as
\begin{equation}
\tau_L^t = A_{\tau} L^{\alpha} {\rm e}^{+F^sL^{d-1}} .
\label{eq:tau}
\end{equation}
The parameters $A_{\tau}$ and $\alpha$ can in principal  be determined
by a fit to the measured tunneling times.

The multicanonical MC algorithm samples configurations with
the weight
\begin{equation}
{\cal P}_L^{mc} (S) \sim\ e^{(\alpha_L^k +\beta_L^k S)}
~~~{\rm for}~~~ S_L^{k} < S \le S_L^{k+1},
\label{eq:pmk}
\end{equation}
instead of sampling with the usual Boltzmann factor
$P_L^B (S) \sim \exp (\beta^c_L S)$ corresponding to the canonical
ensemble.
Here we  partitioned the total action interval $0 \leq S \leq 2  V$
into $k=0,...,N$ (N odd) intervals $I_k=(S^k_L,S_L^{k+1}]$.
The idea of the multicanonical MC algorithm is
to choose intervals $I_k$ and
values of $\beta_L^k$ and $\alpha_L^k$
at the pseudocritical point $\beta_L^c$
in such a way, that
the resulting
multicanonical action density $\cal P$$_L(S)$ has a approximately flat behavior
for values of the action in the interval  $[S_L^{1,max},S_L^{2,max}]$, that
is to say: configurations dominated by the interface are no longer
exponentially suppressed as they are
in the metastable-unstable region of the canonical ensemble.
Physically this can be achieved by choosing the $\beta$-parameters
$\beta_L^k$  such, that the system
gets heated, when its in the ordered state
of the metastable region, cooled
when its in the disordered state, and neither of both if its in the unstable
region.
The parameters $\beta_L^k$
hereby take the form $\beta_L^k=\beta_L^c + \delta \beta_L^k$, where the
coupling constant difference $\delta \beta_L^k$ changes sign as a function
of $S$ and is responsible for the altered dynamics of
the model. The parameters $\alpha_L^k$ are adjusted in such a way that
$\cal P$$_L(S)$ is a steady function of $S$.

In \cite{our} we demonstrated, that when
the double peak distribution $P_L(S)$ can be approximated by a double gaussian,
the multicanonical action density $\cal P$$_L(S)$ can be made arbitrarily
flat by driving a control parameter $r>1$ towards $1$. In this case 
we choose action values  $S^k_L$ with $S^0_L=0$, $S^{N+1}_L=2V$,
$S^1_L=S^{1,max}_L$, $S^{N}_L=S^{2,max}_L$ and in the interval
$[S^{1,max}_L,S^{min}_L )$  action values defined
by the equation $P_L(S^{k}_L)=r^{1-k} P_L(S^{1,max}_L)$ for $k=1,...,N/2$.
An analog procedure is adopted in the interval $(S^{min}_L,S^{2,max}_L]$.
Having defined the action values $S_k$ and corresponding
intervals $I_k$ the setting
\begin{equation}
\beta_L^k\ =\ \cases{\beta_L^c\ {\rm for}\ k=0,N/2,N \cr
\beta_L^c +  \ln (r) / (S_L^{k+1}-S_L^k)\ {\rm for}\ k=1,...,N/2-1  \cr
\beta_L^c -  \ln (r) / (S_L^{k+1}-S_L^k)\ {\rm for}\ k=N/2+1,...,N-1 \cr}
\label{eq:mkx}
\end{equation}
and the recursion
\begin{equation}
\alpha_L^{k+1}\ =\ \alpha_L^k + (\beta_L^k-\beta_L^{k+1}) S_L^{k+1},\
  ~\alpha_L^0\ =\ 0
\label{eq:mky}
\end{equation}
defines the multicanonical ensemble. In accordance with detailed balance,
standard Metropolis and heat bath updating algorithms have been
generalized to the multicanonical situation \cite{our,Bau}. Finally
we obtain the
canonical action density distribution $P_L(S)$
through a reweighting step similar to \cite{FS,BauBer} from the
multicanonical distribution $\cal P$$_L(S)$:
\begin{equation}
P_L(S)\ = e^{ (\beta^c_L - \beta^k_L S
- \alpha_l^k )} {\cal P}_L (S) \ {\rm for} \
S_L^{k} < S \le S_L^{k+1}.
\label{eq:rew}
\end{equation}

As an example we show for our $L=70$ system in Fig. 3  the
multicanonical action density distribution $\cal P$$_{70}(S)$
together with the reweighted distribution $P$$_{70}(S)$.
In practice the appropriate choice of the parameters in eq.~(7)
is obtained by making from the given systems a FSS prediction of the
density distribution $P_L(S)$ for the next larger system.
A second run may then be performed with
optimized parameters. It is our experience that the guess works normally
so well that the second run is barely an improvement as compared with
the first. On the smallest systems standard MC simulation
provides initial data.
Our statistics for this investigation was
$4 \cdot 10^6$ heat bath sweeps per run and lattice size. One sweep
updates each spin of the lattice once.

We define the tunneling time $\tau_L^t$ as the average number of sweeps
needed to get from a configuration with action $S=S^{1,\max}_L$
to a configuration with $S=S_L^{2,\max}$ {\bf and} back.
With our statistics of $4 \cdot 10^6$ sweeps per run
the system tunnels then in total $8\cdot 10^6 / \tau_L^t$
times. Table~1. collects
the measured tunneling times.
For our smaller systems we have also carried out standard heat bath
MC runs at $\beta^c_L$ and the associated tunneling times are also
reported in Table~1. For the larger systems standard MC runs would
not tunnel often enough to allow for a reliable direct calculation of
their tunneling times. This is of course due to the exponential slowing
down of the standard MC simulation.
In Fig. 4 we display on a log to log scale the divergence
of the tunneling times $\tau^t_L$ for the multicanonical MC algorithm
(circles) and the heat bath algorithm (triangles). There
is clearly a different behavior of the two algorithms involved.
While for the multicanonical MC algorithm the increase of the tunneling time
is consistent with a power law, the heat bath algorithm displays an exponentially
fast growing tunneling time. Performing a $\chi^2-$fit
we obtain the
following fits
\begin{eqnarray}
 \tau_L^t  ({\rm multicanonical })& = & 0.73(3) \cdot L^{2.65(2)}
\ ~~{\rm with}~~ \ \chi^2/(n_F-1)=0.96 , \\  \tau_L^t  
({\rm heat \ bath      })& = & 1.46 \cdot L^{2.15} \cdot e^{+0.080 \times L}
\ ~~{\rm with}~~  \ \chi^2/(n_F-1)=1.34 .
\label{eq:fits}
\end{eqnarray}
The quality of the fits as indicated by the $\chi^2$ values ($n_F-1$
denotes the number of degrees of freedom minus one) is reasonable.
In case of the heat bath algorithm we could not reliably determine
the errors from a $3$ parameter fit. The ratio
$R\ =\ \tau_L^t(\rm heat bath) ~/~ \tau_L^t (\rm multicanonical) $
is a direct measure for the relative efficiency of the two algorithms.
Using the fits we extrapolate its value to the $L=100$ system and
estimate a factor $R \approx 500$ for this case. The multicanonical
algorithm approximately slows down like $\sim V^{2.325}$ with respect
to the number of updates per degree of freedom.
This is only slightly worse than the optimal performance $\sim V^2$
which was estimated in \cite{our} based on a random walk picture.
For the heat bath algorithm the inefficiency of the algorithm
prohibits a very accurate estimate of $F^s$ from the behavior of the
tunneling time according to eq.~(6). The fitted value in eq.~(11) is
however close to the determination of the next paragraph (eq.~(15)).

Our multicanonical data allow the so far most precise determination of
the interfacial free energy for the $2d$ 10-state Potts model. For this
purpose we determine maxima and minima of the $P_L (S)$ distributions
by self-consistent straight line fits over suitable $S$-ranges.
Together with the central values of their associated ranges, our
$F^s_L$ values are collected in Table~2. Performing the FSS fit
according to \cite{face}
\begin{equation}
F^s_L\ =\ F^s + {c \over L}
\label{eq:fitfs}
\end{equation}
we obtain consistent  results for lattices of size $L=16,24,34,50,70$ and
$L=100$, as displayed in Fig.~5 (we have $\chi^2/(n_F-1)=0.27$). We 
estimate the infinite volume value of the interfacial free energy to be
\begin{equation}
F^s\ =\ 0.09822 \pm 0.00079         .
\label{eq:infinite}
\end{equation}
This value  may however depend weakly on the analytical form of the
FSS fit \cite{face} and even with our large lattices we may still face
additional systematic errors of a similar order than the quoted statistical
error. Future simulations on even larger lattice sizes might therefore be 
of interest.

  In summary, we have introduced a multicanonical ensemble for the numerical
simulation of first order phase transitions, which eliminates
an exponentially fast increase of the tunneling time between the
ordered and disordered states in the critical region of the system.
This finding is achieved by replacing the usual equilibrium
dynamics of the canonical ensemble, through a new equilibrium dynamics,
where the ordered and disordered states of the system get heated and cooled in
a well controlled way. Thus configurations dominated by the presence of the
interface are enhanced during the simulation.

  The multicanonical MC algorithm gives a general framework for the
numerical studies of first order phase transitions in statistical
mechanics as well as for field theoretic models. From the numerical point
of view the interesting question will be what improvement factors
can be achieved as compared to standard algorithms for certain models
on certain lattice sizes.
We expect the answer to this question to be determined by 
the value of the quantity $Q=F^s_L \times L^{d-1}$, where the
strength of the
first order phase transition is indicated by the
magnitude of $F^s_L$ and the $d$ is the dimensionality of the system.
In the case of the $2d$ 10-state potts model we find at values
of $Q  \sim 10 $ approximately an improvement of two to three orders
of magnitude, while at values of $Q \sim 1$ the improvement is marginal.

An implementation of the multicanonical MC algorithm for non-Abelian gauge
theories is straightforward and we think that future investigations
of the QCD deconfining phase transition will benefit from this. Beyond
first order phase transition, it may well be that multicanonical algorithms
could be of use for other numerical calculations in statistical
mechanics, like estimates of the free energy or spin glass simulations.
\hfill\break \vskip 20pt

\section*{Acknowledgements}
Our simulations were performed on the SCRI cluster of fast RISC
workstations and the Convex $C240$ at the University of Bielefeld.
We would like to thank Nelson Alves for collaboration
in the early phase of this work.
One of us (T.N.) appreciates discussions with
A. M. Ferrenberg and D. P. Landau during his stay at the Center
for Simulational Physics in Athens, Georgia.

\vfill \hfill \eject

\eject

\section*{Tables}

\begin{table}[h]
\centering
\begin{tabular}{||c|c|c||}                    \hline
 $L$ & $\tau_L^t$ multicanonical & $\tau_L^t$ heat bath \\ \hline
 12  &           542 (4)    &          793 (7)      \\ \hline
 12  &             --       &          776 (9)      \\ \hline
 16  &          1147 (10)   &         1988 (23)     \\ \hline
 24  &          3354 (57)   &         9634 (408)    \\ \hline
 34  &          8375 (245)  &        43923 (3151)   \\ \hline
 50  &         23763 (1321) &       270565 (63222)  \\ \hline
 50  &         24932 (1064) &             --        \\ \hline
 70  &         69492 (6383) &             --        \\ \hline
 70  &         62218 (5560) &             --        \\ \hline
100  &        160334 (16252)&             --        \\ \hline
\end{tabular}
\caption{{\em The tunneling times $\tau_L^t$ as a function of the
lattice size $L$ for the multicanonical MC algorithm (second row)
and the heat bath algorithm (third row).
For some lattice sizes we display the results of several simulations,
whose difference lies in slightly different coupling parameters.
}}
\end{table}

\begin{table}[h]
\centering
\begin{tabular}{||c|c|c|c|c|c||}                    \hline
$L$ & $\beta^c_L$ & $S_L^{1,max}$ & $S_L^{min}$ & $S_L^{2,max}$ & $F^s_L$
                                                             \\  \hline
 $12$   & 1.40738 (09) &   116 &   169 &   243 & 0.1071 (06) \\ \hline
 $16$   & 1.41534 (12) &   216 &   309 &   429 & 0.1086 (07) \\ \hline
 $24$   & 1.42100 (08) &   523 &   723 &   978 & 0.1058 (08) \\ \hline
 $34$   & 1.42338 (09) &  1072 &  1466 &  1945 & 0.1039 (13) \\ \hline
 $50$   & 1.42481 (06) &  2358 &  3162 &  4192 & 0.1027 (11) \\ \hline
 $50$   & 1.42469 (06) &  2357 &  3186 &  4190 & 0.1006 (10) \\ \hline
 $70$   & 1.42536 (06) &  4661 &  6257 &  8190 & 0.0983 (20) \\ \hline
 $70$   & 1.42541 (05) &  4660 &  6250 &  8178 & 0.1007 (12) \\ \hline
$100$   & 1.42576 (04) &  9602 & 13060 & 16686 & 0.0986 (18) \\ \hline
$100$   & 1.42577 (04) &  9577 & 13093 & 16711 & 0.0994 (15) \\ \hline
\end{tabular}
\caption{{\em The pseudocritical couplings $\beta^c_L$, interfacial free
energies and locations of the maxima and minima of the action density
distribution, as determined from the multicanonical distributions.}}
\end{table}

\hfill \vfill \eject

\section*{Figure captions}
\vspace{10pt}
\hspace{5pt}

{\bf Fig. 1} \hspace{5pt}
Canonical action density distributions $P_L(s)$ for $L=24$ and $L=100$ lattices
at the pseudocritical couplings $\beta^c_24$ and $\beta^c_100$.
Here $s$ denotes $s=S/(2V)$.
The values of the maxima have been normalized to $1$.
\vspace{10pt}

{\bf Fig. 2} \hspace{5pt}
Action density distributions  $P_L(s)$
for lattices of size $L=16,24,34,50,70$ and $100$
on a logarithmic scale.
The values of the maxima have been normalized to $1$.
\vspace{10pt}

{\bf Fig. 3} \hspace{5pt}
Multicanonical action density distribution
$\cal P$$_{70}(s)$ together with
with its reweighted distribution $P_{70}(s)$.
The normalization of the distributions is chosen arbitrarily such that
the figure looks nice.
\vspace{10pt}

{\bf Fig. 4} \hspace{5pt}
Tunneling times for the  multicanonical MC algorithm
and the heat bath algorithm in a double log scale.
The curves correspond to the fits in eq. (11) and eq. (12).
The dashed part of the curve indicates the extrapolation to the
$L=100$ lattice for the heat bath algorithm.
On the $100$ lattice the systems still tunnels $50$ times
between the metastable states during $4 \times 10^6$ sweeps, when
the multicanonical simulation is used.
\vspace{10pt}

{\bf Fig. 5} \hspace{5pt}
FSS estimate of the interfacial free energy $F^s$. Averages are used
for those lattices for which we have two data sets.
\vspace{10pt}

\end{document}